\documentclass[twocolumn,showpacs,prb]{revtex4}

\usepackage{graphicx}% Include figure files
\usepackage{dcolumn}% Align table columns on decimal point
\usepackage{bm}% bold math

\begin{document}

\title{Quasiparticle Dynamics in the Kondo Lattice Model at Half Filling}

\author{Simon Trebst$^{1,2}$, Hartmut Monien$^{3}$, Axel Grzesik$^{3}$,  and Manfred Sigrist$^{1}$}\affiliation{$^{1}$Theoretische Physik, Eidgen\"ossische Technische Hochschule Z\"urich,
              CH-8093 Zurich, Switzerland\\
              $^{2}$ Computational Laboratory, Eidgen\"ossische Technische Hochschule Z\"urich, 
              CH-8092 Zurich, Switzerland\\
              $^{3}$Physikalisches Institut, Universit\"at Bonn, Nu\ss allee 12, 53115 Bonn, Germany}

\date{\today}

\begin{abstract}
We study spectral properties of quasiparticles in the Kondo lattice model in one and two dimensions
including the coherent quasiparticle dispersions, their spectral weights and the full 
two-quasiparticle spectrum using a cluster expansion scheme.  We investigate the evolution of the quasiparticle band as antiferromagnetic correlations are enhanced towards the
RKKY limit of the model. 
In both the 1D and the 2D model we find that a repulsive interaction between
quasiparticles results in a distinct antibound state above the two-quasiparticle 
continuum. The repulsive interaction is correlated with the emerging antiferromagnetic 
correlations and can therefore be associated with spin fluctuations. 
On the square lattice, the antibound state has an extended s-wave symmetry.
\end{abstract}


\maketitle

% Introduction -------------------------------------------------------------------------
%\section{Introduction}

The Kondo lattice is one of the fundamental microscopic models for the description 
of heavy fermion materials. The basic ingredients are nearly localized $f$-electrons
on every lattice site and itinerant conduction electrons which are non-interacting apart from a
local coupling to the $f$-electrons. In the Kondo lattice model the $f$-electrons form localized spin
degrees of freedom and the coupling is represented as an onsite exchange interaction between
this spin and the conduction electron spin density. This rather simple model 
gives rise to complex many body physics whose
detailed understanding is still far from complete. Generally it is believed, that 
many of the properties of heavy Fermion systems originate from
the interplay between magnetic RKKY interaction among the localized spins 
and the Kondo effect screening these spins \cite{Doniach:77,Coleman:01,Tsunetsugu:97}. 
The former leads to a long-range ordered
antiferromagnetic phase in two and three dimensions 
and the latter to a phase with  short-ranged spin correlations due to the formation of coherent
Kondo spin singlets. There is a quantum phase transition between the two limiting phases upon
changing the parameters of the model \cite{Doniach:77,Coleman:01,Tsunetsugu:97}. 

For the metallic heavy Fermion systems \cite{Si:01,Coleman:01} this quantum phase 
transition  has recently become one of the high-priority issues. In particular, the behavior of the charge 
carriers at the quantum phase transition has gained special interest and questions have been raised
on the character of the  Fermi surface, whether it is ''large'' or ''small'' \cite{Tsunetsugu:97}. 
In the case of a large Fermi surface
both the conduction electrons and the $f$-electrons forming the localized spins are included in the
Fermi volume. The small Fermi surface, in contrast,  consists only of the conduction electrons. 
It seems obvious that in the case of magnetic long-range order the $f$-electrons tend to loose their
mobility and drop out in the Fermi volume count. On the other hand, the generally accepted picture
of the heavy Fermion phase is that the electrons at the Fermi level have strong $ f$-character
such that the $f$-electrons are in the Fermi volume. Consequently a swift change of
the Fermi surface topology is expected at the magnetic quantum phase transition \cite{Coleman:01}.
In a recent experiment Paschen {\em et al.} have indeed observed a characteristic feature in the
evolution of the Hall effect of YbRh$_2$Si$_2$ through the quantum phase transition, consistent
with this picture \cite{Paschen:04}. 

The Kondo insulator which corresponds to the half-filled Kondo lattice with one conduction electron
per localized spin is a special phase where the quantum phase transition can be more
easily discussed, since there are no low-lying quasiparticle degrees of freedom. 
In the so-called strong coupling regime the Kondo effect dominates yielding a
spin liquid phase with a spin and a charge gap in the excitation spectrum. On the other
hand the antiferromagnetically ordered phase in the weak coupling regime has only a charge gap
while the spin sector possesses gapless spin wave modes. 

While under these circumstances the issue of the Fermi surfaces is not of immediate relevance, examining the behavior of the quasiparticle and charge excitations  will still give much insight into 
the fate of the quasiparticle spectrum for the 
system in the vicinity of the quantum phase transition. In particular, the approach from the quantum
disordered side allows us to investigate  the modification of the quasiparticle spectrum which reflects
the gradually extending magnetic correlations in the approach of the quantum phase transition.
It is also worth to analyze how the enhancement of spin fluctuations affects the two particle excitations.
One of the important questions is the spin fluctuation mediated interaction between two
particles, whether they can form bound pairs. This would give insight into the possibility of
unconventional superconductivity in weakly doped Kondo lattice systems.  

Some of these issues have been addressed recently 
by Assaad and coworkers based on Quantum Monte Carlo
methods as we will discuss later \cite{Assaad:99,Feldbacher:02,Assaad:04}. 
Discussion of the behavior of the half-filled Kondo lattice model and the related
periodic Anderson model in a magnetic field examined the behavior of the
quasiparticle spectrum at quantum phase transitions from a different view point
 \cite{Beach:04,Milat:04}. 
In these latter studies the quantum phase transition to magnetic long-range order has been induced by a 
magnetic field which lowers the spin triplet excitations and the modifications of the quasiparticle spectrum discussed. Both studies show that the system always remains in an insulating phase 
\cite{Beach:04,Milat:04}. 

The Hamiltonian of the Kondo lattice model (KLM) is given by
\begin{equation}
  H_{KLM} = -t \sum_{\langle ij \rangle, \sigma} \left( c^{\dagger}_{i\sigma}c^{\phantom{\dagger}}_{j\sigma} +h.c.\right) + J \sum_i { S}_i \cdot { S}^c_i \,,
 \label{Kondo:KLM}
\end{equation}
where ${ S}^c_i = \frac{1}{2}\sum_{\sigma, \sigma'}{ \tau}_{\sigma, \sigma'}c^{\dagger}_{i\sigma}c^{\phantom\dagger}_{i\sigma'}$ are the spin density 
operators of conduction electrons and ${ S_i}$ are localized $f$-spins at site 
$i$ with $\tau_{\sigma, \sigma'}$ being the Pauli matrices.
The magnetic exchange coupling $J$ derived from the periodic Anderson model is
antiferromagnetic ($J>0$) \cite{Schrieffer:66}. 
Furthermore, $t$ denotes the hopping-matrix element which we restrict to nearest-neighbor
hopping only. The Kondo insulating phase is realized for $ t/J \ll 1 $ where $ t/J $ will be our
small expansion parameter. 

The strong-coupling limit  ($ t/J \to 0 $) represents indeed a good starting point for a well-controlled perturbative approach based on a systematic cluster expansion \cite{Gelfand:00}. 
This serves well our goal to study of the evolution of the quasiparticle spectrum in view of the
emerging longer ranged antiferromagnetic correlations as $ t/J $ is gradually increased. 
In the following sections we will discuss the one- and two-dimensional bipartite Kondo lattice model. 
Although the former does not have a quantum phase transition, the antiferromagnetic correlations
increase strongly for growing $ t/J$. In two dimensions, however, a quantum phase transition
is expected for $ t/J \approx 0.68 $  \cite{Assaad:99}.  

% The one-dimensional model ------------------------------------------------------------
\section{The one-dimensional model}
\label{Section_1D}

  % single quasiparticle dispersion / effective mass
\subsection{The quasiparticle spectrum}
We first discuss spectral properties of quasiparticle (hole) 
excitations for the one-dimensional KLM at half-filling. Starting 
from the limes of decoupled singlets on the lattice sites 
for the Kondo insulator $(t/J = 0)$ 
we have calculated high-order strong-coupling expansions in the
hopping strength $t/J$. 
This allows us to block-diagonalize the Hamiltonian by integrating
out spin and charge fluctuations up to a given order and calculate 
effective Hamiltonians for the ground-state at half-filling, and the 
degenerate manifolds with a single or two quasiparticle (hole) excitations
\cite{Gelfand:00,Trebst:00,Zheng:01a}.
Complementary to previous approaches using bond operator techniques 
\cite{Jurecka:01, Feldbacher:02} our strong-coupling expansion treats charge 
and spin fluctuations on the same footing without the need of any mean field 
approximations. Previous numerical work have studied the nature of the 
ground state of the 1D model as
well as the elementary spin, quasiparticle and charge gaps, such as an extensive DMRG
study \cite{Shibata:99} and strong coupling analysis \cite{Shi:95,Zheng:02}. 
Here we present results of a strong coupling analysis for the single and two particle properties.

\begin{figure}[t]
  \includegraphics[width=\columnwidth]{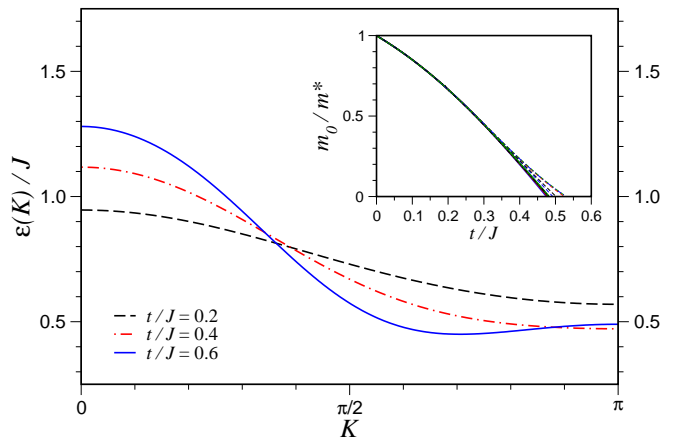}
  \caption{Dispersion of quasiparticles in the 1D Kondo lattice model for 
           various hopping amplitudes $t/J$ obtained by an 11th order strong coupling 
           expansion around the Kondo insulator. The series were extrapolated using 
           optimized perturbation theory.
           The inset shows the reciprocal effective mass. 
           Several Pad\'e and Dlog Pad\'e approximant are shown, as well as parabolic 
           fits to the dispersion obtained by optimized perturbation theory.}
  \label{Dispersion_1D}
\end{figure}

The dispersion of a single quasiparticle excitation shown in Fig.~\ref{Dispersion_1D}
is obtained by a Fourier transformation of the respective effective Hamiltonian
which we have calculated up to 11th order in $t/J$. 
The minimum of the quasiparticle (hole) dispersion is found at momentum $K=\pi$.
With increasing hopping strength the band flattens around its minimum exhibiting
a continuous enhancement of the effective mass. This effect is connected with the
growing ''coherence'' among the $f$-electrons as the localized spins start to be
correlated. This behavior very much resembles what is seen in theories discussing
the heavy Fermion physics in the periodic Anderson model in terms of a renormalized
hybridization of conduction and $ f$-electrons which leads to a minimum of the hole 
excitations at $ K= \pi$ with nearly localized ($f$) character \cite{Rice:85,Coleman:84}. 
Interestingly our calculations show even a
divergent effective mass at $K = \pi$ at a critical value of $ t/J $ beyond which a continuous 
shift of the band minimum away from $K = \pi$ occurs. In the inset of Fig.~\ref{Dispersion_1D} 
we plot the reciprocal effective mass $m_0/m^{*}$
which we have computed from the series for the quasiparticle dispersion.
For various
extrapolation schemes including Pad\'e / Dlog-Pad\'e approximants \cite{DombGreen:74}
and optimized perturbation theory (OPT) \cite{Knetter:04} the effective mass is found
to diverge around $t/J \approx 0.50 \pm 0.02$ independent of the order  of the expansion
(8th to 11th order).
Thus the divergence of the effective mass is obviously not an artifact of the 
strong coupling expansion. Simultaneously for large hopping strength 
quasi long range AF order builds up in the 
one-dimensional system and the spin gap $\Delta_S$ remains finite 
\cite{Tsvelik:94,Shibata:99}.

\begin{figure}[t]
  \includegraphics[width=76mm]{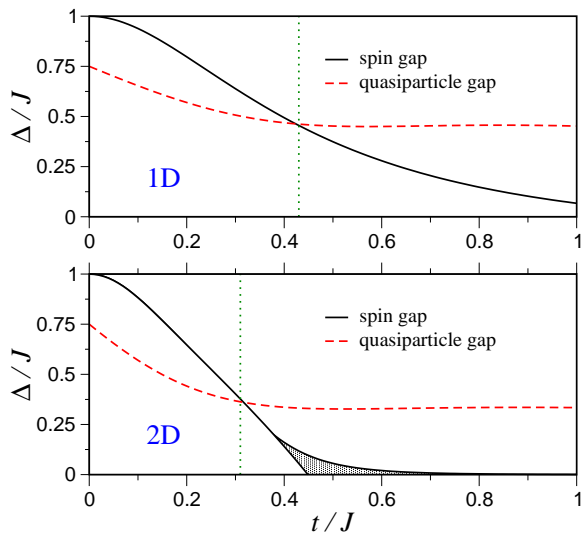}
  \caption{Spin and quasiparticle gap for the Kondo lattice model.
           In one dimension (upper panel) the spin gap stays finite for increasing hopping strength.
           In two dimensions (lower panel) the spin gap vanishes for finite hopping strength, the 
           grey shaded area indicates the variation of the various Pad\'e approximants.
           The quasiparticle gap $\Delta_{QP}$ becomes nearly constant for $t/J > 0.5$ 
           in both dimensions. In 1D (2D) the spin gap crosses the quasiparticle gap 
           $\Delta_S \approx \Delta_{QP}$ around $t/J \approx 0.43 (0.36)$ as indicated by the
           dotted lines. }
  \label{TimeScales}
\end{figure}

The shift of the energy minimum  is connected with the appearance of effective hopping processes 
(obtained through integrating out higher energy configurations) which are non-bipartite, i.e. connected
points on the A (or B) sublattice. To lowest order the dispersion is given by
\begin{equation}
\varepsilon(K) = t \cos K + \frac{t^2}{3J} \cos 2K 
\end{equation}
which yields an effective mass
\begin{equation}
\frac{m_0}{m^*} = 1 + \frac{4t}{3J}
\end{equation}
with $ m_0 = 2/t $. The second term involves the next-nearest neighbor hopping process. 
Such hoppings become increasingly facilitated compared to the hopping between different sublattices, as the antiferromagnetic correlation
grows. This gives a preference to hole transfers between sites of the same sublattice
whose localized spins tend to be parallel. 
Indeed we observe that the appearance of this behavior is correlated by the decrease of
the spin gap below the quasiparticle gap, see Fig.~\ref{TimeScales}. 
As a consequence, the time scale for spin fluctuations increases and the quasiparticle excitation is
dressed by a slowly fluctuating AF spin background.

% The 2-quasiparticle spectrum
\subsection{The two-quasiparticle spectrum}

\begin{figure}[t]
  \includegraphics[width=\columnwidth]{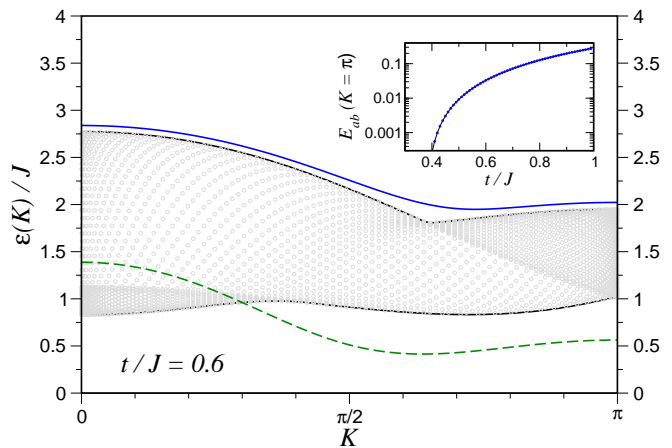}
  \caption{Spectrum of quasiparticles in the 1D Kondo lattice model. 
           For $t/J > 0.4$ an antibound singlet state (solid line) emerges above the 
           continuum (shaded area).
           The single quasiparticle dispersion is given by the dashed line.
           The inset shows the antibinding energy of the singlet antibound state at 
           $K=\pi$. Note the logarithmic scale.}
  \label{TwoQuasiparticleSpectrum_1D}
\end{figure}

We now turn to the spectrum of two-quasiparticle excitations and explicitly address
the question whether two quasiparticles attract or repel each other, thereby forming
a distinct bound or antibound state separated from the two-quasiparticle scattering
continuum. By means of a strong-coupling expansion the two-quasiparticle spectrum
is computed by calculating effective Hamiltonians in the two-particle sector integrating
out order by order spin and charge fluctuations. The cluster expansion then allows us
to determine the {\em exact} Schr\"odinger equation for two quasiparticles in the
thermodynamical limit (up to the order calculated) which we then solve numerically
\cite{Trebst:00,Zheng:01a}.

Here we present results from an expansion  up to 8th order in the hopping strength $t/J$. 
The obtained series for the matrix elements of the effective Hamiltonian have been 
extrapolated by applying the OPT approach \cite{Knetter:04}.
The full spectrum of two quasiparticles is shown in Fig.~\ref{TwoQuasiparticleSpectrum_1D}. 
In addition to the single quasiparticle band (dashed line) there is a continuum of 
scattering states. 
While we do not find any bound states, a singlet antibound state emerges on top of the 
continuum for sufficiently large hopping strength $t/J>0.4$  as illustrated by the solid line
in Fig.~\ref{TwoQuasiparticleSpectrum_1D}. This antibound state starts to split off from
a kink in the upper continuum edge around $K \approx 0.65\pi$. This kink arises for
$t/J > 0.45$ as the minimum of the single quasiparticle band starts to wander away from
$K=\pi$. The singlet antibound state definitely emerges strongly with the enhanced antiferromagnetic correlations.  We find no evidence for a (anti-)bound state in the $S=1$ spin sector.

% The two-dimensional model ------------------------------------------------------------
\section{The two-dimensional model}

For the two-dimensional Kondo lattice model antiferromagnetic spin correlation develops 
with increasing hopping strength much more strongly than in the one-dimensional model
and for a finite critical value of $(t/J)_c$ the system undergoes a quantum phase 
transition from the Kondo insulator, a phase with gaped charge and spin excitations 
and short ranged correlations to an antiferromagnetic state with gapless spin excitations 
and long ranged correlations.  
The critical point was best determined by a quantum Monte Carlo study \cite{Assaad:99} to be 
$(t/J)_c = 0.68 \pm 0.02$. Alternative numerical approaches included
bond-operator mean field calculations \cite{Jurecka:01,Feldbacher:02} and series
expansions \cite{Shi:95,Zheng:02}, which yield similar estimates. 
Here we present results of a strong coupling analysis for the single and two particle properties.

\subsection{The quasiparticle spectrum}

\begin{figure}[t]
  \includegraphics[width=\columnwidth]{./QuasiParticle_2D.eps}
  \caption{Dispersion of quasiparticles in the 2D Kondo lattice model for
           various hopping amplitudes $t/J$ obtained by an 11th order strong coupling
           expansion around the Kondo insulator. The series were extrapolated using
           optimized perturbation theory (OPT).
           The inset shows the reciprocal effective mass. Several Pad\'e and Dlog Pad\'e
           approximante are shown, as well as parabolic fits to the OPT dispersion.}
  \label{Dispersion_2D}
\end{figure}

\begin{figure}[t]
  \includegraphics[width=\columnwidth]{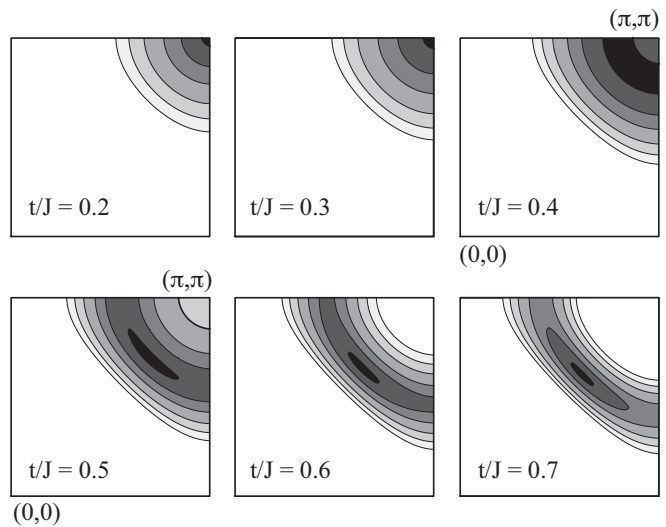}
  \caption{Minimum of the quasiparticle band for the two-dimensional Kondo lattice 
           model, where the grey shading/contour lines reflect the distance from the
           overall minimum (black).
           The panels describe the path from strong Kondo coupling ($t/J = 0.2$, upper 
           left panel) to the antiferromagnetically ordered phase ($t/J = 0.7$, lower 
           right panel). 
           The shift of the band minimum towards the zone center at $K=(\pi/2,\pi/2)$ for
           $t/J>0.375$ indicates the onset of antiferromagnetic order before the system
           undergoes a phase transition at $(t/J)_c = 0.68$. }
  \label{FermiSurface}
\end{figure}

We first discuss spectral properties of single quasiparticle (hole) excitations at half-filling.
The dispersion of the quasiparticle illustrated in Fig.~\ref{Dispersion_2D} has been 
calculated from a strong-coupling expansion around the Kondo insulator up to 11th 
order in the hopping strength $t/J$ summing up some 1,691 cluster diagrams.
Similar to the one-dimensional model, we find the minimum of the dispersion at 
$K=(\pi,\pi)$. For small $t/J<0.3$ our results agree well with a recent series expansion 
study \cite{Zheng:02} and bond operator mean field calculations 
\cite{Jurecka:01,Feldbacher:02} which did not take into account spin fluctuations.
With increasing hopping amplitude we again observe an increasing band width 
accompanied by an increase of the effective quasiparticle mass. The inset in 
Fig.~\ref{Dispersion_2D} shows the reciprocal effective mass $m_0/m^{*}$ of the 
quasiparticle. Analogous to the one-dimensional case we encounter the formation of
a weakly dispersive part of the band close to $ K= (\pi, \pi ) $ which also here is consistent
with the heavy Fermion picture obtained by means of other methods, but is in contrast to
results from a recent quantum Monte Carlo study \cite{Assaad:04}.

For $t/J \approx 0.40 \pm 0.05$ we find that the effective quasiparticle mass diverges, 
and the minimum of the quasiparticle band starts to shift toward the zone center at 
$K=(\pi/2,\pi/2)$ as illustrated in Fig.~\ref{FermiSurface}. 
Similar to the one-dimensional model this shift occurs around the hopping strength where 
the time scales of spin and charge fluctuations become comparable, see the lower 
panel in Fig.~\ref{TimeScales}. In the same way as for the one-dimensional Kondo lattice we
can attribute this behavior to an increased hopping of quasiparticles on the same sublattice.
The quasiparticle dispersion thereby reveals the onset 
of antiferromagnetic spin order well below the transition to the long range ordered state. 

  %The 2-quasiparticle spectrum
\subsection{The two-quasiparticle spectrum}

The singlet antibound state found for the one-dimensional model is the result of 
growing spin fluctuations. In two dimensions the formation of bound or antibound states
due to the effective interaction among the quasiparticles 
has large freedom as we discuss in the Appendix. 
To study this aspect for the two-dimensional model we have calculated the effective 
two-quasiparticle Hamiltonian, integrating out spin and charge fluctuations up to 8th order
in the hopping strength $t/J$.

\begin{figure}[t]
  \includegraphics[width=\columnwidth]{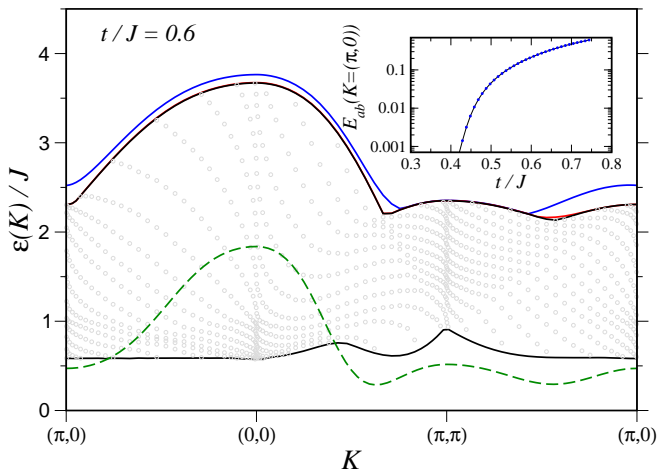}
  \caption{Spectrum of quasiparticles in the 2D Kondo lattice model.
           Close to the parameter regime where the effective quasiparticle mass diverges
           and the time scales of spin and charge fluctuations become comparable, a 
           singlet antibound state with extended $s$-wave symmetry emerges above the 
           continuum. 
           The inset shows the antibinding energy of the singlet antibound state at 
           $K=(\pi,0)$. Note the logarithmic scale.}
  \label{TwoQuasiparticleSpectrum_2D}
\end{figure}

Fig. \ref{TwoQuasiparticleSpectrum_2D} illustrates the full spectrum of quasiparticles
in the 2D Kondo lattice model close to the transition to the AF phase. 
Above the quasiparticle continuum a singlet antibound state (solid line) is found with 
a dispersion bearing some resemblance to the single quasiparticle band (dashed line).
The antibound state only slowly separates from the continuum starting to emerge around
$t/J \gtrsim 0.4$. The antibinding energy at $K=(\pi,0)$ is plotted in the inset of 
Fig. \ref{TwoQuasiparticleSpectrum_2D}.
The circumstance that the singlet antibound state occurs only at finite hopping
strength $t/J$ close to the phase transition to the antiferromagnetically ordered state
provides further evidence that the repulsive interaction originates from 
antiferromagnetic spin correlations.

Performing a symmetry analysis as detailed in an Appendix we 
find that the singlet antibound state has an extended $s$-wave symmetry, since the state
can be mapped onto the irreducible representation $A_1$ of the $C_{4v}$ point symmetry 
group of the two-dimensional square lattice.

% Fermi surfaces
\section{Conduction electrons and heavy quasiparticles}

\begin{figure}[t]
  \includegraphics[width=\columnwidth]{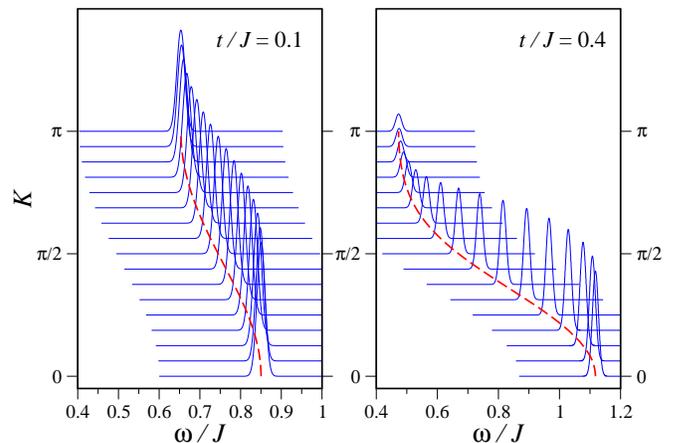}
  \caption{Spectral weight of a quasiparticle (hole) excitation for the one-dimensional
           Kondo lattice model. The $\delta$-peaks are broadened by a Gaussian. 
           The dashed lines indicate the dispersion of the quasiparticle.
           As the hopping amplitude increases a major part of the spectral weight 
           shifts towards the band maximum at $K=0$.}
  \label{SpectralWeights_1D}
\end{figure}

\begin{figure}[t]
  \includegraphics[width=\columnwidth]{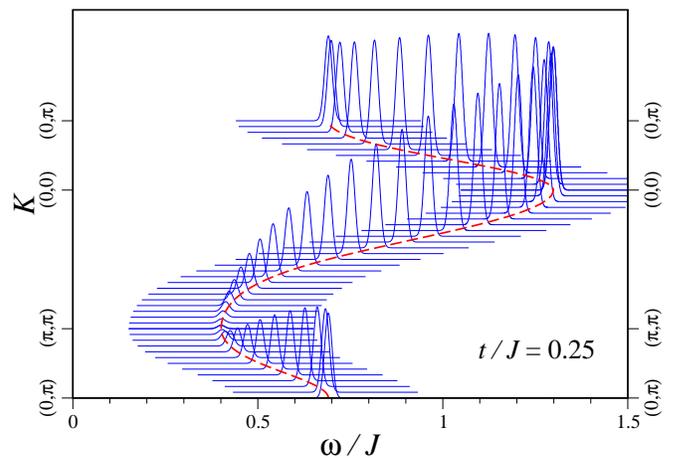}
  \caption{Spectral weight of a single quasiparticle in the two-dimensional Kondo lattice 
           model. The $\delta$-peaks are broadened by a Gaussian. 
           Already for small hopping strength $t/J=0.25$  the spectral weight concentrates at
           the maximum of the quasiparticle band (dashed line) at $K=(0,0)$. }
  \label{SpectralWeights_2D}
\end{figure}

We now consider in more detail the quasiparticle spectrum and relate our
result with other theories of the heavy Fermion state, which we mentioned already
previously. For this purpose we turn to 
 the spectral weight $Z_{QP}(K)$ of the quasiparticle 
excitation by expanding the coherent part of the dynamic structure factor
\begin{eqnarray}
  S(q, \omega) & = & \int \frac{dt}{2\pi} e^{-i\omega t} 
                    \sum_{r,s} e^{iqr} \langle c^{\dagger}_s(0,0) c_s(r,t) \rangle
                    \nonumber \\
  & = & Z_{QP}(K)\delta(\omega - \epsilon_{QP}(K)) + {\text{incoherent}} \;.
\end{eqnarray}
In the strong coupling limit the weight is constant for all $ K $. With growing $ t/J $ a
weight redistribution is observed giving a larger weight on the dispersing part 
for small $ K $ and a pronounced reduction of weight in the nearly flat band region towards
$ K = \pi $ and $ K=(\pi,\pi) $ in one or two dimensions as shown in Figs.~\ref{SpectralWeights_1D} 
and \ref{SpectralWeights_2D}, respectively. 
The same trend has  been observed also in Refs.~\cite{Capponi:01,Feldbacher:02,Assaad:04} 
using Quantum Monte Carlo techniques. 
These calculations suggest that the lost weight is absorbed into a
''shadow band'' being formed due to the enhanced antiferromagnetic correlations. 
Within the strong coupling expansion the incoherent part is understood as a continuum
of the single quasiparticle combined with an independent spin-1 excitation. Out of this
continuum the ''shadow band'' is emerging. We omit a more detailed consideration of
the shadow band here and concentrate on the coherent part. 

\begin{figure}[t]
  \includegraphics[width=\columnwidth]{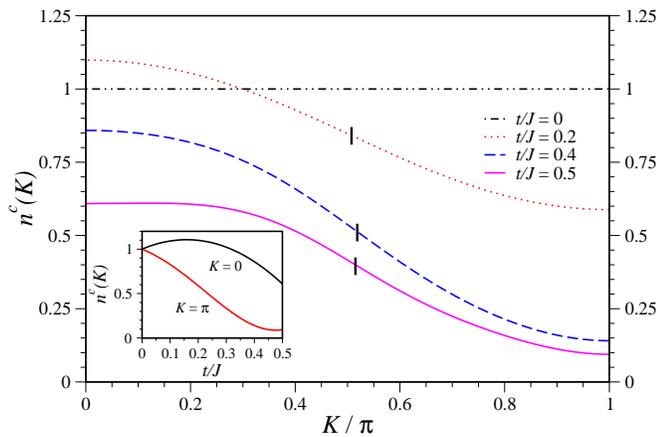}
  \caption{Integrated weight $n^c(K)$ for varying hopping strength $t/J$ in the one-dimensional 
                  Kondo lattice model. The inflection points which are marked by the solid bars indicate 
                  the position of the conduction electron Fermi surface.
                  The inset shows the particle density at $K=0$ and $K=\pi$ versus the hopping strength. }
  \label{IntegratedWeight_1D}
\end{figure}

\begin{figure}[t]
  \includegraphics[width=\columnwidth]{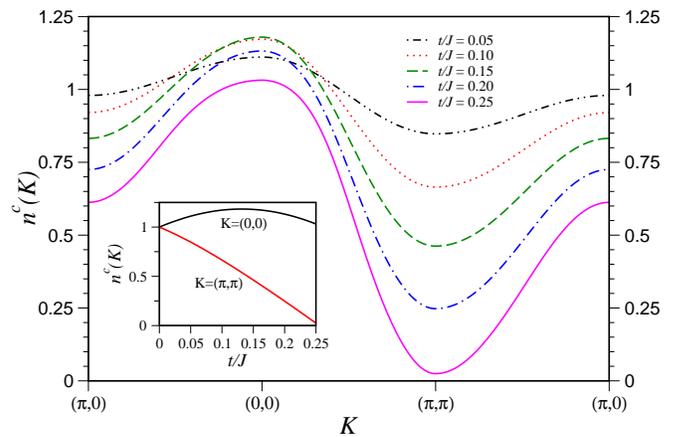}
  \caption{Integrated weight $n^c(K)$ for the two-dimensional Kondo lattice model. 
                  The inset shows the particle density at $K=0$ and $K=\pi$ versus the hopping strength.}
  \label{IntegratedWeight_2D}
\end{figure}

The integrated weight under the coherent quasiparticle peak in $S(q, \omega)$ provides
$ n^c_{K} = \langle c_{Ks}^{\dag} c_{Ks}^{\phantom\dag} \rangle $. 
In Fig.~\ref{IntegratedWeight_1D} we show $ n^c(K) $ for the
one dimensional system. Obviously the large-$K$ part is depleted of the conduction
electron contribution and a relative increase of the density appears in the part of small
$ K $ where the energy scale is given by the hopping matrix element of the conduction 
electrons. The same feature occurs in two dimensions as shown in Fig.\ref{IntegratedWeight_2D}. 
The parts with high weight are bound to eventually become the genuine conduction electron bands. 
Note that there is an overall drop of quasiparticle weight due to many body effects. 

It is interesting to compare our finding with the numerical DMRG results by Ueda and coworkers for the 
one dimensional Kondo lattice with a finite doping \cite{Ueda:94}. 
In their data for $ n^c(K) $ show a large value for small momenta and a pronounced 
drop at $ K $ corresponding to Fermi momentum of the conduction electrons. For higher 
momenta $ n^c(K) $ is much smaller. 
For finite small doping into this band 
the real Fermi level lies in the weakly dispersing part, yielding the ''heavy''
quasiparticles corresponding to a large Fermi surface in the sense of quasiparticle count. 

\begin{figure}[b]
  \includegraphics[width=\columnwidth]{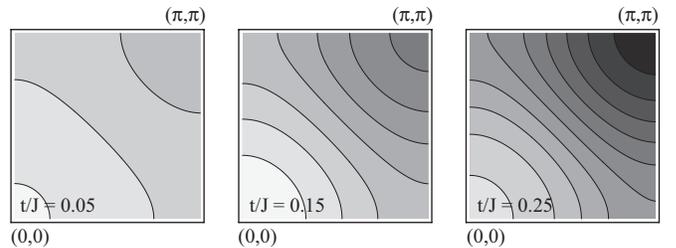}
  \caption{Contour plot of the integrated weight $n^c(K)$ for the two-dimensional Kondo lattice model.}
  \label{Fermicontours}
\end{figure}

We can also follow the mean field and Gutzwiller-type of discussion of the
periodic Anderson or Coqblin-Schrieffer model which give for half-filling:
\begin{equation}
n^c(K) = \frac{1}{2} \left (1 + \frac{\gamma (K)}{\sqrt{\gamma(K)^2 + \alpha^2}}  \right)
\end{equation}
with $ \gamma(K) = 2t \cos K $ (1D) and $ \gamma(K)= 2t (\cos K_x + \cos K_y) $ \cite{Rice:85,Coleman:84,Read:84}. The parameter
$ \alpha $ is a measure for the effective hybridization between the conduction and $ f$-electrons,
which has different renormalizations depending on the corresponding
mean field treatment and the model. 
This form closely resembles our cluster expansion form. We may read out the position of
the conduction electron ''Fermi surface'' defined roughly as the inflection points of $ n^c(K) $.
This yields the point close to $ K = \pi /2 $ in one dimension (Fig.~\ref{IntegratedWeight_1D}) and essentially the square-shaped form in two dimensions (Fig.~\ref{Fermicontours}). 
These are the ''small'' Fermi surfaces which
 are the relevant ones in the RKKY-picture, 
where the $f$-electrons are considered as entirely localized spins. 

The suppression of $ n^c(K) $ occurs in the region of the Brillouin zone
where the band is getting more and more dispersionless, i.e. this heavy quasiparticle part which
may be considered as having  ''nearly localized $f$-electron'' character of the quasiparticles. 
This aspect looks 
surprising in view of the fact the Kondo lattice model in a rigorous manner does not allow for
$f$-electron charge fluctuations. Thus this behavior is here entirely mediated via the entanglement of
conduction and $f$-electron spin. From this viewpoint we may ask when these quasiparticles cease to
exist. While our  cluster expansion method is definitely unable to give any information  beyond 
the quantum phase transition inside the antiferromagnetic 
phase, one may still guess on the fate of the heavy quasiparticles. While the antiferromagnetic
order will suppress the fluctuations of the localized spins, still there are strong quantum fluctuations
leaving space for the entanglement of the $f$- and conduction electrons. Thus, although it is
difficult to show rigorously within our perturbative scheme, we may speculated that the heavy quasiparticles remain in their place even in the magnetically ordered phase for a certain range beyond 
the quantum phase transition. The heigher weight conduction electron-like quasiparticles together with the shadow band form then gapped quasiparticle band in the reduced Brillouin zone. This would
suggest rather a gradual disappearing of the heavy quasiparticles even in the ordered phase.

% Conclusions --------------------------------------------------------------------------

\section{Conclusions}
In conclusion, we have studied the dynamics of quasiparticles in the one- and 
two-dimensional Kondo lattice model at half filling. 
In the regime of strong Kondo coupling we find typical heavy fermion behavior. The
effective quasiparticle mass gets larger as the mobility of the quasiparticle increases
and spectral weight is shifted towards the quasiparticle band gap.
As the system is driven towards (quasi)long-ranged antiferromagnetic order
the shift of quasipartical weigths signals qualitatively the same effects as 
suggested for the metallic system, the change of the Fermi surface topology. 
Based on our results we may pose the question whether this change be really abrupt.

Furthermore it is interesting to discuss the effect of the growing spin fluctuation on the two-quasiparticle spectrum. Would there be a tendency towards bound pair formation as we expect from RVB-like
systems? Although both the Kondo insulator as well as the RVB system possess short-ranged
spin singlet correlations, the Kondo insulator does not show any sign of pair formation. On the opposite,
antibound states appear in the two-quasiparticle spectrum.

% Acknowledgments ----------------------------------------------------------------------

\begin{acknowledgments}
We  thank  F.~Assaad, I.~Milat and H. Tsunetsugu for many helpful discussions.
This study has been financially supported by the Swiss National Science Foundation and
the NCCR MaNEP.
\end{acknowledgments}

% Appendix -----------------------------------------------------------------------------
\appendix

\section{Symmetry projection of two-particle (anti)bound states}
\label{SymmetryProjection}

To compute the two-quasiparticle speactrum we consider the symmetric 
two-quasiparticle Schr\"odinger equation \cite{Trebst:00}
\begin{eqnarray}
\left(E - E_0 - E_1({K,q})\right)f({ K,q}) \quad = 
\nonumber \\
\frac{1}{N} \sum_{ q'} f({ K,q'})
\left(\, \sum_{{ a},{ a'}} V({ a},{ a'})                          
- E_1({ K,q}) \right) \;,
\label{eq:TwoParticle}
\end{eqnarray}
where $E_0$ is the ground state energy, $E_1({ K,q})$ the combined energy 
of two scattering quasiparticles and $V({ a},{ a'})$ are the irreducible matrix 
elements of the calculated two-quasiparticle effective Hamiltonian. 
The scattering amplitude is denoted as $f({ K,q})$. 
The numerical solution of Eq.~(\ref{eq:TwoParticle}) allows to compute the 
two-quasiparticle continuum and all bound and antibound states.

To determine the symmetry of the antibound state we project the irreducible 
matrix elements of the effective two-quasiparticle Hamiltonian, $V({ a},{ a'})$,
onto the irreducible representations $\gamma$ of the point group $C_{4v}$
\begin{equation}
  V({ a},{ a'}) \longrightarrow
  F_{ a}^{(\gamma)}({ q}) V({ a},{ a'}) F_{ a'}^{(\gamma)}({ q'}) \;,
\end{equation}
where the projection $F_{ a}^{(\gamma)}({ q})$  is given by
\begin{equation}
  F_{ a}^{(\gamma)}({ q}) = \frac{1}{|C_{4v}|} \sum_{g \in C_{4v}}
  \chi_g^{(\gamma)} \cos({ q} \cdot g{ a}) \;.
\end{equation}
We subsequently solve the projected two-quasiparticle Schr\"odinger equation
(\ref{eq:TwoParticle}) and thereby identify onto which representation the antibound
state can be mapped.

\begin{table}
\begin{tabular}{c|rrrrr}
 $C_{4v}$  & $E$ & $C_2$ & 2$C_4$ & 2$\sigma_v$ & 2$\sigma'_v$ \\
 \hline
 $A_1$     & 1 & 1 & 1 & 1 & 1 \\
 $A_2$     & 1 & 1 & 1 & -1&-1 \\
 $B_1$     & 1 & 1 &-1 & 1 &-1 \\
 $B_2$     & 1 & 1 &-1 &-1 & 1 \\
 $E$       & 2 &-2 & 0 & 0 & 0 \\
\end{tabular}
\caption[]{Characters of irreducible representations of the point group $C_{4v}$}
\end{table}

% References
\bibliography{kondo}

\end{document}